\documentclass{ifacconf}
\usepackage{times}
\usepackage{amsmath}
\usepackage{amssymb}
\usepackage{mathtools}
\usepackage{xcolor}
\usepackage{kotex}

\usepackage{graphicx}
\usepackage{algorithm}
\usepackage{algorithmic}
\usepackage{comment}
\usepackage{adjustbox}
\usepackage{romannum}
\usepackage[inkscapeformat=png]{svg}
\usepackage{multirow}
\usepackage{tabularx}
\usepackage{siunitx}
\usepackage{tikz}
\usepackage{pgfplots}
\usepackage{grffile}
\usepackage{graphicx}
\usepackage{natbib}

\newcommand{\cl}{\mathsf{cl}}
\newcommand{\YY}{\mathcal{Y}}
\newcommand{\UU}{\mathcal{U}}
\newcommand{\R}{\ensuremath{{\mathbb R}}}
\newcommand{\N}{\ensuremath{{\mathbb N}}}

\begin{document}
\begin{frontmatter}

\title{ARX-Implementation of encrypted nonlinear dynamic controllers using observer form\thanksref{footnoteinfo}} 

\thanks[footnoteinfo]{This work was supported by the National Research Foundation of Korea (NRF) grant funded by the Korea government (MSIT) (No. RS-2024-00353032).}

\author[snu]{Deuksun Hong} 
\author[snu]{Donghyeon Song} 
\author[snutech]{Mingyu Jeong}
\author[snutech]{Junsoo Kim} 

\address[snu]{ASRI and Department of Electrical and Computer Engineering, \\
   Seoul National University, Seoul, 08826, Korea \\(e-mail: \{dshong, dhsong\}@cdsl.kr)}
\address[snutech]{Department of Electrical and Information Engineering, Seoul National University of Science and Technology, Seoul, 01811, Korea\\(e-mail: jeongmingyu@cdslst.kr, junsookim@seoultech.ac.kr)}

\begin{abstract} 
While computation-enabled cryptosystems applied to control systems have improved security and privacy, a major issue is that the number of recursive operations on encrypted data is limited to a finite number of times in most cases, especially where fast computation is required. To allow for nonlinear dynamic control under this constraint, a method for representing a state-space system model as an auto-regressive model with exogenous inputs (ARX model) is proposed. With the input as well as the output of the plant encrypted and transmitted to the controller, the reformulated ARX form can compute each output using only a finite number of operations, from its several previous inputs and outputs. Existence of a stable observer for the controller is a key condition for the proposed representation.
The representation replaces the controller with an observer form and applies a method similar to finite-impulse-response approximation. It is verified that the approximation error and its effect can be made arbitrarily small by an appropriate choice of a parameter, under stability of the observer and the closed-loop system. Simulation results demonstrate the effectiveness of the proposed method.
\end{abstract}


\end{frontmatter}

\section{Introduction}
As networked control systems become more prevalent, the threat of cyberattacks and the intrusion of unauthorized access have become major problems \citep{Teixeira2015Secure, Ding2021Secure}. Homomorphic encryption offers a powerful solution for enhancing confidentiality. The application of homomorphic encryption to feedback systems is referred to as encrypted control \citep{kogiso2015cdc}. This approach allows a controller to operate directly over encrypted signals without decryption, as proposed in foundational works such as \cite{ Kim2016Encrypting, Farokhi2017Secure}. In this standard architecture, the sensing side of the plant encrypts its measurements with a public key and sends them to the controller. The controller performs its calculations on this encrypted data and sends an encrypted control signal back. The actuator stage of the plant then decrypts this signal with the secret key and applies it to the system.

However, applying homomorphic encryption to dynamic controllers presents a fundamental challenge due to the limitations on the number of recursive operations. In most homomorphic encryption schemes, the number of operations on a ciphertext is limited because accumulated errors eventually cause decryption failure or overflow \citep{CKKS17}. This restriction makes it difficult to implement dynamic controllers, which inherently require recursive updates of the state variable. Since recursive updates of the encrypted state lead to unbounded growth of accumulated error, directly implementing dynamic controllers may be infeasible for an infinite time horizon.

To address this recursion issue, intensive research has been conducted for linear systems \citep{SchulzeDarup2021,Kim2022ARC}. Early attempts considered periodically resetting the controller state \citep{Murguia2020Secure} or assumed re-encryption of the controller state \citep{kogiso2015cdc}. Although such implementations bypass the error growth problem by refreshing the ciphertext, it may incur further communication costs or admit performance degradation.
Accordingly, several techniques have been proposed for linear systems, by converting state matrices to integer forms to avoid error growth \citep{Cheon2018CDC, Kim2023Dynamic} or by reformulating the controller into a linear auto-regressive with exogenous (ARX) model using a finite amount of input-output data \citep{Teranishi2024, Lee2025Encrypted}. Utilizing ``re-encryption of the controller output,'' these can be practical alternatives instead of the state re-encryption, as the controller output is supposed to be decrypted for actuation. Despite these advances in linear encrypted control, research on nonlinear dynamic controllers remains limited.

Since sustaining nonlinear operations indefinitely on encrypted data is generally impossible without using bootstrapping \citep{gentry09}, which is computationally expensive, we propose that representing a state-space nonlinear dynamic operation as an ARX model can be a promising direction.
Assuming that the controller output can be re-encrypted and re-used for the operation and the function, an ARX model (approximated to a polynomial function) as a function of several past inputs and outputs of the system will be able to continue encrypted operations over time without the error overflow problem, using a finite number of multiplication and addition operations for each output.

This paper proposes a method to reformulate a nonlinear dynamic controller as an ARX model to continue its encrypted operation an unlimited number of times. Our key idea is to represent the controller to an observer form and apply a method similar to finite-impulse-response approximation.
Although the given controller may be an unstable system, its observer typically substitutes a portion of the state by a function of its output, so that the represented observer becomes stable with respect to its state, while it can yield the same state and output trajectory with the same initial state used.
Then, a finite-impulse-response approximation like method for stable systems can be applied;
by letting the next state of the system be computed from its several past inputs and outputs instead of the state recursion, the controller can be represented as an ARX form.
It is proposed that an approximation error introduced for the representation can be made arbitrarily small by increasing the number of past inputs and outputs used for the computation, thanks to the stability of the observer.
And, it is verified that the effect of such perturbation in the closed-loop system can also be made arbitrarily small, under closed-loop stability.

This paper is organized as follows. Section~\ref{sec:pf_arx} formulates the problem of representing a dynamic controller as an ARX model. In Section~\ref{sec:pf_arx3},  the method is proposed and the main result is presented.
Linear system case is discussed in
Section~\ref{sec:pf_arx4}.
Section~\ref{sec:pf_arx6} shows the effectiveness of the method through simulation results, and Section~\ref{sec:pf_arx7} concludes the paper.

{\it Notation:}
The set of real numbers and positive integers are denoted by $\R$ and $\N$, respectively.
Let $\|\cdot \|$ denote the (induced) {infinity} norm for vectors and matrices.
A continuous function $\gamma(s)$ is of class-$\mathcal{K}$ if it is strictly increasing and $\gamma(0)=0$.
A function $\beta(s,t)$ is of class-$\mathcal{KL}$,
if $\beta(s,t_0)$ is of class-$\mathcal{K}$ for each fixed $t_0\ge0$ and $\beta(s_0,t)$ is decreasing for each fixed $s_0\ge0$ and $\lim_{t\to\infty}\beta(s_0,t)= 0$. Let $f^N$ denote the $N$-times composition of a function $f$.
Let $0$ denote the zero vector or matrix of appropriate dimension, and
$I$ the identity matrix.

\section{Problem Formulation}\label{sec:pf_arx}
Consider a discrete-time nonlinear plant
\begin{equation}\label{eq:plant}
\begin{aligned}
x_p(t{+}1)&=f_p\big(x_p(t),\,u(t)\big),\quad t=0,1,2,\ldots \\ y(t)&=h_p\big(x_p(t)\big), \, x_p(0)=x_{p,0}
\end{aligned}
\end{equation}
where $x_p(t)\in\R^{n_p}$ is the state with the initial value $x_{p,0} \in \R^{n_p}$, $u(t)\in\R^{n_u}$ is the input, and $y(t)\in\R^{n_y}$ is the output. Let a discrete-time dynamic controller have been designed as
\begin{equation}\label{eq:controller}
\begin{aligned}
x_c(t{+}1)&=f_c\big(x_c(t),\,y(t)\big)+{e_c(t)} \\ u(t)&=h_c\big(x_c(t)\big), \,x_c(0)=x_{c,0}
\end{aligned}
\end{equation}
where $x_c(t)\in\R^{n_c}$ is the state with the initial value $x_{c,0} \in \R^{n_c}$ and $e_c(t)\in\R^{n_c}$ is the perturbation.
The perturbation $e_c(t)$ will indicate the performance error caused by the proposed reformulation compared to the controller $f_c$.
The case $e_c(t)\equiv0$ is referred to as the nominal case, for which the trajectories are denoted as $\{\bar{x}_c(t),\bar{x}_p(t),\bar{u}(t),\bar{y}(t)\}$, respectively. They are assumed to be  bounded by a constant $M$ as 
\begin{equation}\label{eq:bound}
	\left\|\begin{bmatrix}
		\bar x_c(t),
		\bar x_p(t),\bar u(t),\bar y(t)
	\end{bmatrix}^\top \right\| \le M,\quad \forall t\ge 0.
\end{equation}

The nominal trajectories are assumed to be stable with respect to the perturbation $e_c(t)$;
there is a class-$\mathcal{K}$ function $\gamma$ such that
\begin{equation}\label{eq:cl_assumption}
\left\| \begin{bmatrix}
	x_c(t)-\bar x_c(t)\\
	x_p(t) - \bar x_p(t)\\
	u(t) - \bar u(t)\\
	y(t) - \bar y(t)
\end{bmatrix} \right\|
\le \gamma \left( \max_{0\le \tau< t} \| e_c(\tau) \|\right),\quad \forall t\ge0.
\end{equation}

To run the dynamic controller \eqref{eq:controller} over encrypted data (using a fixed times of multiplication for each output), we propose how to represent the controller \eqref{eq:controller} in an auto-regressive model with exogenous inputs (ARX).
With a parameter $N\in\N$ indicating the length of the time-horizon, let us define
\begin{equation*}
	\begin{aligned}
		\mathcal Y_N(t)&:=[y(t),~\dots,~y(t-N+1)]^\top\in \R^{Nn_y},\\
		\mathcal U_N(t)&:=[u(t),~\dots,~u(t-N+1)]^\top\in \R^{Nn_u},
	\end{aligned}
\end{equation*}
for $t\ge N-1$.
Then, a class of systems in the form
\begin{equation}
	u(t) = G_N(\mathcal{Y}_N(t-1), \mathcal{U}_N(t-1)),
\end{equation}
with a static function $G_N$, is called the ARX model of order $N$. In particular, the control law
\begin{equation}\label{eq:ARXC}
	u(t)=
	\begin{cases}
		\bar{u}(t), & t\in[0,\,N),\\[2pt]
		G_N\big(\mathcal Y_N(t-1),\,\mathcal U_N(t-1)\big), & t\in[N,\,\infty).
	\end{cases}
\end{equation}
is referred to as the ARX controller (ARXC) of order $N$,
which computes the state $x_c(t)$ using the given function $f_c$ while $t<N$, and switches to use the function $G_N$ from $t\ge N$.

Our objective is to find a function $G_N$ such that the performance of \eqref{eq:ARXC} is equivalent to that of the given controller \eqref{eq:controller} with an arbitrarily small error. This goal is formalized as follows.
\begin{prob}\label{prob:arxc_performance}
    Given $\epsilon>0$, construct an ARXC such that 
    \begin{equation}\label{eq:goal}
        \| {x_p(t)} - \bar{x}_p(t) \| \le \epsilon,~~~\forall t \ge0,
        \end{equation}
    where $x_p$ is the plant state in the closed-loop of \eqref{eq:plant} and \eqref{eq:ARXC}.
    \hfill$\square$
\end{prob}

Considering an ARXC re-written as the form \eqref{eq:controller} with the perturbation $e_c(t)$, the assumption
 \eqref{eq:cl_assumption} will let us to aim for
\begin{equation}\label{eq:delta}
	\|e_c(t)\|\le \gamma^{-1}(\epsilon)=:\delta,\quad \forall t\ge0
\end{equation}
so that \eqref{eq:goal} is achieved.

The next section will propose that if there is an (stable) observer form for \eqref{eq:controller},
then it will directly allow an ARX implementation.

\section{Main Result}\label{sec:pf_arx3}

The proposed ARX-reformulation utilizes a method similar to finite-impulse-response approximation, which is applicable for stable systems.
To describe the idea, let us temporarily suppose that
the controller \eqref{eq:controller} is contractive (stable) itself;
that is, suppose that
there exists a class-$\mathcal{KL}$ function $\beta_{\text{temp}}$ such that
\begin{multline}\label{eq:contractivity}
	\|f_c^t(x_{c,0},\{y(\tau)\}_{\tau=0}^{t-1})-f_c^t(x'_{c,0},\{y(\tau)\}_{\tau=0}^{t-1}) \|
	\\\le \beta_{\text{temp}}(\|x_{c,0}-x'_{c,0}\|,t),\quad 
	\forall t\ge0
\end{multline}
holds for any initial states $x_{c,0}$ and $x_{c,0}'$, sharing the same input sequence $\{y(\tau)\}$ bounded by $M$.
If so, similarly to the finite-impulse-response approximation for stable linear systems, it will let the state $x_c$ be computed without state recursion, as
\begin{align*}
	x_c^{\text{temp}}(t) = f_c^N(0, \YY_{N}(t-1)),\quad u(t) = h_c(x_c^{\text{temp}}(t))
\end{align*}
with a parameter $N$, which is obviously an ARX form.
Then, $x_c^{\text{temp}}$ would obey \eqref{eq:controller} with the perturbation $e_c(t)$ determined as
$$ e_c(t) = x_c^{\text{temp}}(t+1)-f_c(x_c^{\text{temp}}(t),y(t)). $$
This error would be bounded under the contractivity \eqref{eq:contractivity}, as
\begin{align*}
\|e_c(t)\| &=\|f_c^N(0,\YY_{N}(t)) -f_c(x_c^{\text{temp}}(t),y(t))\|\\ 
&= \|f_c^N(0,\YY_{N}(t)) - f_c^{N}(f_c(0,y(t-N)),\YY_{N}(t))\|\\
&\le \beta_{\text{temp}}(M,N),
\end{align*}
provided that
$$\|f_c(0,y(t-N))\|\le M\quad\text{and}\quad \|\YY_N(t)\|\le M. $$
It follows that choosing $N$ such that $\beta_{\text{temp}}(M,N)\le \delta$ would achieve the goals \eqref{eq:delta} and \eqref{eq:goal}.
However, this method would  not be applicable for an unstable controller.

Then, our idea is to replace the controller \eqref{eq:controller} by an observer form which is stable in terms of its state and apply the above method.
We first assume
the existence of a stable observer form.

\begin{assum}\label{asm:observer_stability}
	\begin{subequations}\label{eq:asm2}
	A continuous map $f_o(x_c,y,u)$ exists such that
	\begin{equation}
		f_o( x_c, y, h_c( x_c)) = f_c( x_c,y),\quad \forall x_c\in\R^{n_c},\forall y\in\R^{n_y},
	\end{equation}
	and  with some class-$\mathcal{KL}$ function $\beta$,
	\begin{multline}
\|f_o^t(x_c,\{y(\tau),u(\tau)\}_{\tau=0}^{t-1})-f_o^t(x'_c,\{y(\tau),u(\tau)\}_{\tau=0}^{t-1})\|\\\le \beta(\|x_c-x'_c\|,t),\quad\forall x_c\in\R^{n_c},\forall x'_c\in\R^{n_c}
	\end{multline}
\end{subequations}
 holds for any trajectories $y(\tau)$ and $u(\tau)$ bounded by $M+\epsilon$.\hfill$\square$
\end{assum}

\begin{rem}
The bound $M+\epsilon$ considers a slight deviation for the nominal trajectories of $\{y(t),u(t)\}$ due to the perturbation $e_c(t)$ in \eqref{eq:controller}.
A global observer is assumed for simplicity, but an observer on a local domain can be considered in practice.\hfill$\square$
\end{rem}

\begin{rem}
	The existence of $f_o$ indeed means the existence of a stable observer;
	under
	Assumption~\ref{asm:observer_stability}, the state observer, as
	\begin{equation*}
			\hat x_c(t+1)=f_o\left(\hat x_c(t),\,y(t),\,u(t)\right),\quad 
			u(t) = h_c( x_c(t))
	\end{equation*}
	receiving the input $y(t)$ and the output $u(t)$ of the controller \eqref{eq:controller},
	will yield a correct estimate as
	$$ \|\hat x_c(t) - x_c(t)\| \le \beta(\| \hat x_c(0)-  x_c(0)\|,t) $$
when $e_c(t)\equiv0$, thanks to the properties \eqref{eq:asm2}.\hfill$\square$
	\end{rem}

Now, we proceed to construct an ARXC using the  function $f_o$ for the controller.
Recall that the nominal trajectory $ x_c(t)=\bar x_c(t)$ of the controller \eqref{eq:controller} (with $e_c(t)\equiv0$) is supposed to obey
\begin{align*}
\bar x_c(t) &= f_c^N(\bar x_c(t-N), \{\bar y(\tau)\}_{\tau=t-N}^{t-1} )\\
&=	f_o^N(\bar x_c(t-N), \{\bar y(\tau),\bar u(\tau)\}_{\tau=t-N}^{t-1})\quad \text{for~ $t\ge N$}
\end{align*}
where the state term $\bar x_c$ in $f_o^N$ will be ignorable as $N$ increases, under the observer stability.
This lets us to compute $x_c(t)$ by
\begin{align}\label{eq:proposed}
	\begin{split}
		x_c(t) &= f_o^N(0,\YY_N(t-1),\UU_N(t-1))\quad\text{for~ $t\ge N$}\\
		u(t) &= h_c(x_c(t))
	\end{split} 
\end{align}
with $N\in\N$ being a parameter.
It is clearly an ARXC for $t\ge N$, with the function $G_N$ in \eqref{eq:ARXC} found as $G_N = h_c\circ f_o^N$. The proposed dynamics for $x_c(t)$ can be identified with the form \eqref{eq:controller} with the perturbation term $e_c(t)$ determined by\footnote{Recall that $f_c(x_c,y)=f_o(x_c,y,h_c(x_c))=f_o(x_c,y,u)$.} 
\begin{align}
	e_c(t) &= x_c(t+1)-f_c(x_c(t),y(t))\label{eq:e_c}\\
	&= f_o^N(0,\YY_N(t),\UU_N(t)) - f_o^{N+1}(0,\YY_{N+1}(t),\UU_{N+1}(t))\notag
\end{align}
for $t\ge N$, and $e_c(t)= 0$ for $ t < N$.

Regarding the performance of the proposed ARXC and the error caused by ignoring $x_c(t-N)$ in the computation,
we claim that the performance error can be made arbitrarily small by increasing the parameter $N$.
Depending on $\epsilon$ and $\delta$ given from Problem~\ref{prob:arxc_performance} and \eqref{eq:delta} (which can be arbitrarily small), indicating a desired upper-bound for the performance error,
the parameter $N$ is proposed to be chosen to satisfy
\begin{equation}\label{eq:N}
	\beta\left(\max_{\|[y,u]^\top\|\le M+\epsilon}{\|f_o(0,y,u)\|},N\right) \le \delta,
\end{equation}
which always exists with $f_o$ being continuous on a compact set.

Finally, the following theorem states the main result.
\begin{thm}\label{thm:main}
	Under Assumption~\ref{asm:observer_stability},
	consider the closed-loop  of the plant \eqref{eq:plant} and the proposed ARXC \eqref{eq:proposed}.
	Given $\epsilon>0$, with the parameter $N$ satisfying \eqref{eq:N}, it guarantees that \eqref{eq:goal} holds.\hfill$\square$
	\end{thm}	
	{\it Proof:}
	Considering the condition \eqref{eq:delta} with the perturbation $e_c(t)$ determined by \eqref{eq:e_c},
	we show that
	\begin{equation}\label{eq:pf}
		\|e_c(t)\|\le\delta,\quad \|y(t)-\bar y(t)\|\le \epsilon,\quad \|u(t)-\bar u(t)\|\le \epsilon
	\end{equation}
	for all $t\ge0$.
For all $t<N$, \eqref{eq:pf} is clearly true with
$e_c(t)=0$, $y(t)=\bar y(t)$, and $u(t)=\bar u(t)$.
	Suppose that \eqref{eq:pf} is true for all $t< \tau$ with some $\tau\ge N$.
	Observe from \eqref{eq:e_c} that
	\begin{align}\label{eq:pf2}
		\|e_c(\tau)\|&=\|
		f_o^N(0,\YY_N(\tau),\UU_N(\tau))\notag\\
		&\,\,\,\,-
		f_o^{N}(f_o(0,y(\tau-N),u(\tau-N)),\YY_N(\tau),\UU_N(\tau))\|\notag\\
		&\le \beta(\|f_o(0,y(\tau-N),u(\tau-N))\|,N),
	\end{align}
	where the function $\beta$ is given by Assumption~\ref{asm:observer_stability}.
	Suppose that \eqref{eq:pf} is true for all $t< \tau$ with some $\tau\ge N$; that is,
    $$
      \|e_c(k)\|\le\delta,\,
      \|y(k)-\bar y(k)\|\le\epsilon,\,
      \|u(k)-\bar u(k)\|\le\epsilon,
      \quad k<\tau.
    $$
    Then, by \eqref{eq:cl_assumption} with $t=\tau$,
    $$
      \left\|
        \begin{bmatrix}
          y(\tau)-\bar y(\tau)\\[2pt]
          u(\tau)-\bar u(\tau)
        \end{bmatrix}
      \right\|
      \le \gamma\!\left(\max_{0\le k<\tau}\|e_c(k)\|\right)
      \le \gamma(\delta)=\epsilon,
    $$
    and hence
    $$
      \|y(k)-\bar y(k)\|\le\epsilon,\quad
      \|u(k)-\bar u(k)\|\le\epsilon,
      \quad k\le\tau.
    $$
    Combining this with the nominal bound \eqref{eq:bound} yields
    $$
      \|y(k)\|\le M+\epsilon,\quad
      \|u(k)\|\le M+\epsilon,
      \quad k=\tau-N,\dots,\tau.
    $$
    Therefore, $\{y(k),u(k)\}_{k=\tau-N}^{\tau}$ satisfies
    the boundedness condition required in Assumption~\ref{asm:observer_stability},
    and in particular
    $$
      \|f_o(0,y(\tau-N),u(\tau-N))\|
      \le \max_{\|[y,u]^\top\|\le M+\epsilon}\|f_o(0,y,u)\|.
    $$
    Thus, \eqref{eq:pf2} together with \eqref{eq:N} ensures that
    $\|e_c(\tau)\|\le \delta$.
	Thanks to the closed-loop stability condition \eqref{eq:cl_assumption}, it follows that
	$$\left\|\begin{bmatrix}
	y(\tau)-\bar y(\tau)\\
	u(\tau)-\bar u(\tau)
	\end{bmatrix}\right\|\le \gamma(\delta)=\epsilon, $$
	so that \eqref{eq:pf} is true for $t=\tau$.
	Therefore, \eqref{eq:pf} is true for all $t\ge0$ by the induction principle, and the proof is completed.
	\hfill$\blacksquare$
	
	Finally, the implications of Theorem~\ref{thm:main} are discussed in the context of computation-enabled cryptosystems, such as homomorphic encryption schemes.
	At each time $t$,
	the controller \eqref{eq:proposed} uses the previous inputs $\YY_N(t-1)$ and the outputs $\UU_N(t-1)$ to compute the current control input $u(t)$.
	Rather than storing an internal state variable for recursive computation, each new computation starts directly from the newly received input and output data.
	Thus, if $u(t)$ and $y(t)$ are provided as ``fresh ciphertexts'' that have not undergone prior computation, the required number of functional operations (typically additions and multiplications) remains fixed according to the structure of $f_o^N$.
	This enables continuous encrypted control, even when the allowable number of operations on each encrypted value is limited, assuming that the controller output $u(t)$ is re-encrypted and transmitted to the controller.

\begin{rem}\label{rem:observer_based_controller}
	The method will be applicable for ``observer-based controllers'' directly.
	If the controller \eqref{eq:controller} is given as the form
	\begin{equation*}
		x_c(t+1)=f_{\text{obs}}\left(x_c(t),y(t),u(t)\right),\quad u(t) = h_c(x_c(t))
	\end{equation*}
	which ensures that $$\|x_c(t)-x_p(t)\|\le \beta_{\text{obs}}(\|x_p(0)-x_c(0)\|,t)$$ with some class-$\mathcal{KL}$ function $\beta_{\text{obs}}$,
	one can easily verify that $f_{\text{obs}}=f_o$ and $\beta_{\text{obs}}=\beta_o$ satisfy the condition \eqref{eq:asm2}.\hfill$\square$
\end{rem}

\section{Linear System Case}\label{sec:pf_arx4}
This section shows how
the method is applied to linear systems, and describes how the parameters $\{f_o,N\}$ can be chosen depending on the bound $\epsilon$.
Let the plant \eqref{eq:plant}
take the form of
\begin{equation*}
    \begin{split}        
    x_p(t+1) &= Ax_p(t)+Bu(t) \\
    y(t) &= Cx_p(t),\quad x_p(0)=x_{p,0}
    \end{split}
\end{equation*}
and let the controller \eqref{eq:controller} be given as
\begin{equation*}
    \begin{split}
        x_c(t+1) &= Fx_c(t) + Gy(t) + e_c(t), \\
        u(t)&= Hx_c(t), \quad x_c(0) = x_{c,0}.
    \end{split} 
\end{equation*}
We write the the closed-loop with $x = [x_p^\top, x_c^\top]^\top$ at once, by
\begin{align}\label{eq:cl}
	\begin{split}
		x(t+1) &= \begin{bmatrix}
			A  & BH \\
			GC & F
		\end{bmatrix}x(t) + \begin{bmatrix}
			0\\
			I
		\end{bmatrix} e_c(t),~ x(0)\!=\!\begin{bmatrix}
		x_{p,0}\\x_{c,0}
		\end{bmatrix}\!=:\!x_0\\
		&=: A_\cl x(t) + B_\cl e_c(t).
	\end{split}
\end{align}
The matrix $A_\cl$ is assumed to be (Schur) stable, so that
$$\|A_\cl^t\|\le M_\cl \lambda_\cl^t,\quad\forall t\ge0$$
with some $M_\cl>0$ and $0<\lambda_\cl<1$.
The bound $M$ in \eqref{eq:bound} for the nominal trajectories can be found such that
$$ M \ge \max\{\|C\|,\|H\|,1\}M_\cl\|x_0\|. $$
The bound for the error caused by the perturbation $e_c(t)$ will be
$$ \|x(t)-\bar x(t)\| \le \left\|\sum_{\tau=0}^{t-1}A_\cl^{t-1-\tau}B_\cl \right\| \max_{0\le\tau<t}\|e_c(\tau)\| $$
so that $\{\gamma,\delta\}$ in \eqref{eq:cl_assumption} and \eqref{eq:delta} are found by linear functions, as
\begin{align*}
	\gamma =
	\max\{\|C\|,\|H\|,1\}\frac{M_\cl}{1-\lambda_\cl},\quad \delta = \gamma^{-1}\epsilon.
\end{align*}

Assumption~\ref{asm:observer_stability} is reduced to the condition that the pair $(F,H)$ of the controller is observable (detectable);
there exists $R\in\R^{n_c\times n_u}$ such that $F-RH$ is stable, and $$\|(F-RH)^t\|\le M_o\lambda_o^t,\quad\forall t\ge0 $$
with some $M_o>0$ and $0<\lambda_o<1$.
The map $f_o$ is found as
$$ f_o(x_c,y,u) = (F-RH)x_c + Gy + Ru $$
with which the function $\beta$ is found as
$ \beta(s,t) = M_o\lambda_o^t s$.
An ARXC for $t\ge N$, with the parameter $N$, is obtained as
\begin{equation}\label{eq:controller_linear}
	x_c(t) = 
		\sum_{\tau=0}^{N-1} (F-RH)^\tau (G y(t-\tau) + R u(t-\tau) ),\quad \text{for}~t\ge N,
\end{equation}
and \eqref{eq:N} suggests that the parameter $N$ be chosen to satisfy
\begin{multline}\label{eq:N_linear}
(||G||+||R||)(M+\epsilon)M_o\lambda_o^N\le \gamma^{-1}\epsilon\\
\iff N\ge \frac{1}{\log \lambda_o}\log\left(\frac{\gamma^{-1}\epsilon}{(||G||+||R||)(M+\epsilon)M_o}\right).
\end{multline}

Under these parameter choice, we have the following corollary.

\begin{cor}
	Assuming that $F-RH$ is stable,
	the ARXC \eqref{eq:controller_linear} with the parameter $N$ satisfying \eqref{eq:N_linear} ensures that \eqref{eq:goal} holds.\hfill$\square$
\end{cor}

To provide a less conservative parameter design for linear systems,
we calculate the performance error using the $z$-transformation,
instead of relying on Theorem~\ref{thm:main}.
Note that
the error $e_c(t)$
in the closed-loop \eqref{eq:cl} is determined from \eqref{eq:e_c}, as
\begin{align*}
	e_c(t) &= - (F-RH)^N\begin{bmatrix}
		GC&RH
	\end{bmatrix}x(t-N)\\
	&=: -\Delta_Nx(t-N)
\end{align*}
where we have $x(\tau)=0$ for $\tau<0$.
Let $X(z)$ and $\bar X(z)$ denote the (unilateral) $z$-transform\footnote{Define $X(z):= \sum_{\tau=0}^{\infty} x(\tau)/z^\tau $. } of $x(t)$ and $\bar x(t):=[\bar x_p(t)^\top,\bar x_c(t)^\top]^\top$, respectively.
Then, \eqref{eq:cl} turns into
\begin{align*}
zX(z) - zx_0 &= A_\cl X(z) - B_\cl \Delta_N\frac{X(z)}{z^N}\\
z\bar X(z) - zx_0 &= A_\cl \bar X(z)
\end{align*}
so the $z$-transform $E(z)$ of $e(t):= x(t)-\bar x(t)$ is obtained by\footnote{Note that $P_N^{-1}-P^{-1}= P_N^{-1}(P-P_N)P^{-1}$.}
\begin{align}\label{eq:E}
E(z) 
&= \left(\left(zI-A_\cl + \frac{B_\cl \Delta_N}{z^N}\right)^{-1} - (zI-A_\cl)^{-1}\right)zx_0\notag\\
&=:(P_N(z)^{-1} - P(z)^{-1})zx_0\notag \\
&= P_N(z)^{-1}\left(\frac{-B_\cl\Delta_N}{z^{N-1}}\right)P(z)^{-1}x_0.
\end{align}
Given that the matrix $A_\cl$ is stable and $\lim_{N\rightarrow \infty}\Delta_N=0$,
note that the unit circle $|z|=1$ lies within the region of convergence when $N$ is sufficiently large.
This allows us to consider
\begin{equation}
	\|x(t)-\bar x(t)\| = \frac{1}{2\pi}\left\|\int_{0}^{2\pi}
	E(e^{j\omega})e^{j\omega n}d\omega\right\|\le
	\max_{\omega\in\R}\|E(e^{j\omega})\|
\end{equation}
which will become arbitrarily small as $N$ increases.
As a result, we have the following proposition.
\begin{prop}
Assume that $F-RH$ is stable. For any $\epsilon>0$, there exists $N'\in\N$ such that
for any $N\ge N'$,
the function $E(z)$ is stable and $\max_{\omega\in\R}\|E(e^{j\omega})\|\le\epsilon$, so that \eqref{eq:goal} holds.\hfill$\square$
\end{prop}

{\it Proof:}
Note in \eqref{eq:E} that the poles of $E(z)$ are the roots of
\begin{align*}
	\det(z^{N+1}-A_\cl z^N + B_\cl \Delta_N)&=0\\
	\det (zI-A_\cl)&=0
\end{align*}
and zeros.
As $N$ increases, these roots
approach the zeros and the eigenvalues of $A_\cl$ arbitrarily closely, which ensures the stability of $E(z)$ due to the stability of $A_\cl$.
The fact that
$$\lim_{N\rightarrow\infty}\max_{\omega\in\R}\|E(e^{j\omega})\|=0$$
directly follows, because
$$\max_{\omega\in\R}\|P_N(e^{j\omega})^{-1}\|<\infty\quad\text{and}\quad\max_{\omega\in\R}\|P(e^{j\omega})^{-1}\|<\infty $$
when $N$ is sufficiently large, and $\|B_\cl\Delta_N x_0\|$ tends to zero as the parameter $N$ tends to infinity.
This completes the proof.\hfill$\blacksquare$

\begin{rem}
	As discussed in Remark~\ref{rem:observer_based_controller}, the method is directly applicable for observer-based controllers, without requiring the observability of  $(F,H)$ to find $R$.
	If the controller is given as
	\begin{align*}
		x_c(t+1) &= (A-LC)x_c(t) + Ly(t) + Bu(t)\\
		u(t) &= Kx_c(t),
	\end{align*}
	given that $A-LC$ is stable, the matrices $\{F-RH,G,R\}$ in \eqref{eq:controller_linear} can be replaced by $\{A-LC,L,B\}$, respectively.\hfill$\square$
\end{rem}

\begin{rem}
The case of ``deadbeat observer'' is notable, as it occurs when all the eigenvalues of $F-RH$ are zero.
Since $(F-RH)^{n_c}=0$ for this case,
having $N=n_c$ ensures that $e_c(t)=0$ for all $t\ge0$, as investigated in \citep{Teranishi2024,Lee2025Encrypted}, which exploits the observability of $(F,H)$.\hfill$\square$
\end{rem}

\section{Simulation Results}\label{sec:pf_arx6}
We consider the single-link flexible joint robot plant \citep{IbrirXieSu2005}.
The discrete-time nonlinear plant model is given by
\begin{align*}
    x_p(t+1) &= Ax_p(t) + f(x_p(t)) + Bu(t), \\
    y(t) &= Cx_p(t)
\end{align*}
where $x_p\in\R^4$, $u(t)\in\R$,  $y=[y_1,y_2]^\top\in\R^2$, and
\begin{align*}
	&A = \begin{bmatrix}
		1 & 0.01 & 0 & 0 \\
		-0.486 & 0.9875 & 0.486 & 0\\
		0 & 0 & 1 & 0.01\\
		0.195 & 0 & -0.195 & 1
	\end{bmatrix}, \quad
	C = \begin{bmatrix}
		1 & 0 & 0 & 0\\
		0 & 1 & 0 & 0
	\end{bmatrix}, \\
	&f([x_1,x_2,x_3,x_4]^\top) = \begin{bmatrix}
		0 \\
		0 \\
		0 \\
		-0.0333\sin(x_{3})
	\end{bmatrix}, \quad\!
	B = \begin{bmatrix}
		0 \\
		0.216 \\
		0 \\
		0
	\end{bmatrix}.
\end{align*}

An observer-based nonlinear controller is designed as
\begin{equation}\label{eq:simulation_controller}
\begin{aligned}
    x_c(t+1) &= Ax_c(t) + f(x_c(t))+Bu(t) +L(y(t)\!-\!Cx_c(t)), \\
    u(t) &= Kx_c(t),
\end{aligned}
\end{equation}
with
\begin{align*}
    K &= \begin{bmatrix}
            -20.4547 & -6.0923 & 14.3017 & -2.1379
         \end{bmatrix}, \\
    L &= \begin{bmatrix}
            0.9994 & 0.0047 \\
            -0.5037 & 1.2477 \\
            -0.0492 &0.5631 \\
            0.1986 & 0.4025
         \end{bmatrix}.
\end{align*}

With this choice of $L$, the nonlinear observer map $(A - LC)x + f(x)$ is stable at the origin.
Since the controller \eqref{eq:simulation_controller} takes a form of a stable observer by itself,
as noted in Remark~\ref{rem:observer_based_controller},
we apply the finite-impulse-response approximation method on \eqref{eq:simulation_controller} without an additional design.

The simulation setup is as follows. The initial states are set to $x_p(0)=[-2.0, 0, 0, 0]^\top$ and $x_c(0)=[0, 0, 0, 0]^\top$. 
The simulation runs for a duration of $T=300$ steps.
To reflect quantization effects in implementation, all controller and observer parameters are represented with four significant digits, with trailing zeros omitted in their notation.
In the scenario, the system initially operates with the given controller and switches to the proposed ARXC at the time step $t=20$.

\begin{figure}[t]
    \centering
    \input{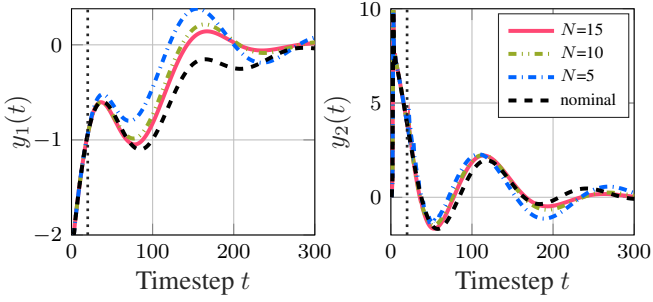}
    \caption{Plant output trajectories with respect to the order $N$.}
    \label{fig:output_comparison}
\end{figure}

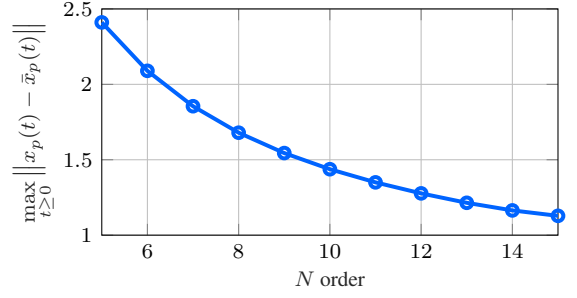
\begin{figure}[t]
    \centering
    \definecolor{mycolor1}{RGB}{0,100,255}

\begin{tikzpicture}

\begin{axis}[
width=6cm,
height=3.0cm,
at={(4cm,0cm)},
scale only axis,
xmin=5,
xmax=15,
xlabel style={font=\color{white!15!black}, yshift = 5pt},
xlabel={\footnotesize $N$ order},
xticklabel style={font=\small},
yticklabel style={font=\small},
ymin=1,
ymax=2.5,
ylabel style={font=\color{white!15!black}, yshift = -10pt},
ylabel={\footnotesize $\max\limits_{t \geq 0}\left\|x_{p}^{}(t) - \bar{x}_{p}(t)\right\|$},
axis background/.style={fill=white},
xmajorgrids,
ymajorgrids
]
\addplot [color=mycolor1, line width=1.5pt, mark=o, mark options={solid, mycolor1}, forget plot]
  table[row sep=crcr]{%
5	2.41072397775405\\
6	2.0896058976006\\
7	1.85500739896295\\
8	1.67937838691542\\
9	1.54430944054007\\
10	1.43728012815873\\
11	1.35007474746009\\
12	1.27709537067943\\
13	1.21457549940386\\
14	1.16414456118948\\
15	1.12818058955329\\
};
\end{axis}

\end{tikzpicture}%
    \caption{Maximum error with respect to the order $N$.}
    \label{fig:norm_difference}
\end{figure}

Figs.~\ref{fig:output_comparison} and~\ref{fig:norm_difference} illustrate the effect of increasing the order $N$ from $5$ to $15$. 
As $N$ increases, the plant trajectories under the ARXC closely approach those of the nominal closed-loop system. 
Fig.~\ref{fig:output_comparison} displays the plant output responses, while Fig.~\ref{fig:norm_difference} illustrates the maximum norm of the plant state difference between the nominal controller and the ARXC with respect to the order $N$.
As proposed, the effect of the perturbation and error for reformulating the controller becomes negligibly small as the parameter $N$ increases, under stability.

\section{Conclusion}\label{sec:pf_arx7}
We have introduced a method for reformulating nonlinear dynamic controllers into ARX models, to enable encrypted operations to be continued without recursive multiplications. By replacing the given controller by an observer form and applying a method similar to finite-impulse-response approximation, the state recursion operation has been replaced by a static function of several past inputs and output. As a consequence, the encrypted dynamic operation becomes realizable through output re-encryption. Each output can then be computed using a finite number of operations, without relying on state recursion.
Future work will aim to further realize encrypted dynamic control using the ARX implementation.
The effects of quantization and polynomial approximation for the ARX models should be taken into account.
Relaxing the observer–existence assumption will also be of interest, as it would accommodate a broader class of nonlinear systems.

\bibliography{ifacconf}             
\end{document}